\documentclass[aps,pra,twocolumn,groupedaddress, showpacs]{revtex4}

 \usepackage{graphicx}
 \usepackage{amssymb}
 \usepackage{epstopdf}
\begin{document}

\title{Electron electric dipole moment experiment using \\ electric-field quantized slow cesium atoms} 

\author{Jason M. Amini}
 \email{JAmini@PhysicsJazz.info}
  \altaffiliation[Also at]{Physics Department, University of California at Berkeley, Berkeley CA 94720.} 
 \altaffiliation[Present Address ]{MS 847.00, National Institute of Standards and Technology, 325 Broadway, Boulder, Colorado 80305.}
  \author{Charles T. Munger Jr.}
 \email{Charles@SLAC.stanford.edu}
 \altaffiliation[Also at ]{Mailstop 61, SLAC, 2575 Sand Hill Road, Menlo Park, CA 94025}
\author{Harvey Gould}
 \email{HAGould@lbl.gov}
 \homepage{http://homepage.mac.com/gould137}
\affiliation{Mail Stop 71-259, Lawrence Berkeley National Laboratory, Berkeley, CA 94720}

\date{\today}

\begin{abstract}
A proof-of-principle electron electric dipole moment (e-EDM) experiment using slow cesium atoms, nulled magnetic fields, and electric field quantization has been performed. With the ambient magnetic fields seen by the atoms reduced to less than 200 pT, an electric field of 6 MV/m lifts the degeneracy between states of unequal $|m_F|$ and, along with the low ($\approx$ 3 m/s) velocity, suppresses the systematic effect from the motional magnetic field. The low velocity and small residual magnetic field have made it possible to induce transitions between states and to perform state preparation, analysis, and detection in regions free of applied static magnetic and electric fields. This experiment demonstrates techniques that may be used to improve the e-EDM limit by two orders of magnitude, but it is not in itself a sensitive e-EDM search, mostly due to limitations of the laser system. 
 \end{abstract}

\pacs{32.60.+i, 32.10.Dk, 14.60.Cd, 32.80.Pj}

\maketitle

\section{INTRODUCTION}
\subsection{Electron Electric Dipole Moments and Extensions of the Standard Model}
A permanent electron electric dipole moment (e-EDM) in an eigenstate of angular momentum exists only if parity (P) and time-reversal (T) are violated, where T violation is equivalent to charge-parity (CP) violation. No EDM of any particle or system has yet been observed: all known CP violation (in the decays of the B and K$^0$ systems) is consistent with the Standard Model's Cabbibo-Kobayashi-Maskawa (CKM) mechanism. The CKM mechanism directly affects only the quark sector and the CKM-generated e-EDM is extremely small. It is estimated \cite{bern91, bern91erra, pospelov05} to be about $10 ^{-10} to 10^{-5}$ (depending upon assumptions about the number of neutrino generations and their masses) of the current e-EDM experimental limit of $2.6 \times 10^{-48}$ C-m ($1.6 \times 10^{-27}$ e-cm) \cite{regan02} (see also \cite{hudson02, abdullah90, murthy89}) --- and beyond the sensitivity of presently planned experiments.

The observation of an e-EDM would signify a new, non-CKM source of CP violation \cite{bern91, bern91erra, wolfenstein04, pospelov05}. New, non-CKM sources of CP violation, that affect leptons directly and that can give rise to a potentially measurable e-EDM, are contained in extensions of the Standard Model. A non-CKM source of CP violation is thought to be necessary to generate the observed excess of matter over antimatter in the universe \cite{sakharov67}.

Potentially observable e-EDMs \cite{bern91, bern91erra, barr93, pospelov05} are predicted by Supersymmetry \cite{abel01}, Multi-Higgs Models, Left-Right Symmetric Models, Lepton Flavor-Changing Models, and Technicolor Models \cite{applequist04}. Split Supersymmetry \cite{arkani-hamed05, chang05, giudice06} predicts an e-EDM in a range from the present experimental limit to a few orders of magnitude smaller.  Improving the present e-EDM limit would place constraints on Standard Model extensions and possibly on current models of neutrino physics \cite{mohapatra05}. Even in the absence of new particle discoveries at accelerators, observing an e-EDM would prove that there was new physics beyond the Standard Model, 

\subsection{Electron EDM Experiments}
Laboratory e-EDM experiments search for a difference in energy between an electron aligned and anti-aligned with an external electric field. (Alternatively a change in the rate of precession of the electron spin may be sought.) High atomic number paramagnetic atoms and molecules provide test systems of zero net charge and can enhance the sensitivity to an e-EDM. The calculated enhancement factor $R$ for the cesium ground state is $114\pm15$ \cite{sandars66, johnson86}. Other atoms of interest, Tl and Fr, have enhancement factors of -585 and 910 respectively \cite{liu92, byrnes99}. Because the interpretation of the e-EDM measurement does not depend on subtracting out CKM effects, the error in the enhancement factor does not need to be small. 

A cesium e-EDM experiment detects an EDM as a shift in the energy between different  ($z$ components of total angular momentum) $m_F$ hyperfine sublevels that is linear in an applied electric field. To avoid a false positive, non-EDM effects that produce shifts that are likewise linear in the applied electric field must be suppressed. Because both the electron's dipole moments (magnetic and electric) are proportional to the electron spin, magnetic fields that change synchronously with the electric field can mimic an e-EDM. Examples include magnetic fields from leakage currents across electric field structures; magnetic fields set up by relays used for electric field reversal; and for moving atoms and molecules, the magnetic field from the Lorentz transform of the applied electric field, the so-called motional magnetic field.

\begin{figure} [h]
\includegraphics [scale = 0.35] {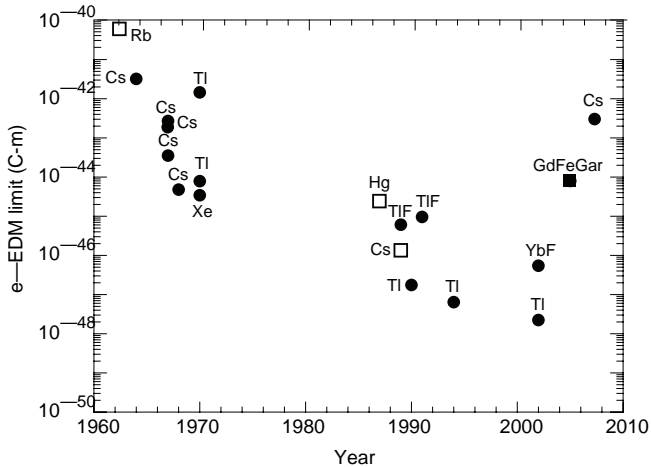}
\caption{\label{history}
Experimental upper limits to the e-EDM 1962 --- 2007. Atomic and molecular beam experiments are shown as filled circles, cell experiments as open squares and solid state experiments as filled squares. The atom, molecule, or solid used is indicated.}
\end{figure}
Since 1964, improvements in the control of systematic effects have allowed the limit on the e-EDM to be lowered by about six orders of magnitude, as shown in Fig.~\ref{history}. Most experiments used thermal beams of atoms \cite{sandars64,angel67,stein67,stein67err,carrico68,stein69,weisskopf68,carrico70,gould70, player70,abdullah90,commins94,regan02}, but thermal beams of molecules \cite{cho89,cho91,hudson02}, atoms confined in buffer-gas filled cells \cite{ensberg62, ensberg67, lamoreaux87,murthy89}, and recently solids \cite{heidenreich05} have also been used. 
For thermal beams of atoms, the most important systematic effect is caused by  the motional magnetic field \cite{sandars64}. 

The motional magnetic field $\textbf{B}_{\rm mot}$, seen by a neutral atom moving with velocity $\textbf{v}$ through an electric field $\textbf{E}$ is (S.I. units)
\begin{equation} \label{mot}
\mathbf{B}_{\rm mot} = \mathbf{v \times  E}/c^2.
\end{equation}
Here $c$ is the speed of light. When a static magnetic field $\textbf{B}_{\rm 0}$, such as may be used to lift the degeneracy between $m_F$ levels, is also present, misalignment between $\textbf{E}$ and $\textbf{B}_{\rm 0}$ causes a component of $\textbf{B}_{\rm mot}$ to lie along $\textbf{B}_{\rm 0}$. This component is linear in $\textbf{E}$ and hence mimics an EDM.

To suppress the motional magnetic field effect, thermal Cs and Tl atomic beam experiments used velocity cancellation from colinear beams traveling in opposite directions \cite{angel67,abdullah90,commins94,regan02}, or alignment of $\textbf{E}$ and $\textbf{B}_{\rm 0}$ with low-enhancement-factor alkali atoms serving as the alignment magnetometer \cite{stein67, stein67err, weisskopf68,stein69,gould70,regan02}, or both \cite{regan02}.  After six orders of magnitude of improvement in suppressing the motional magnetic field effect, these techniques may have reached a practical limit, as is evidenced by a slowing in the rate of improvement in the e-EDM limit in Fig.~\ref{history}.

\begin{figure}
\includegraphics {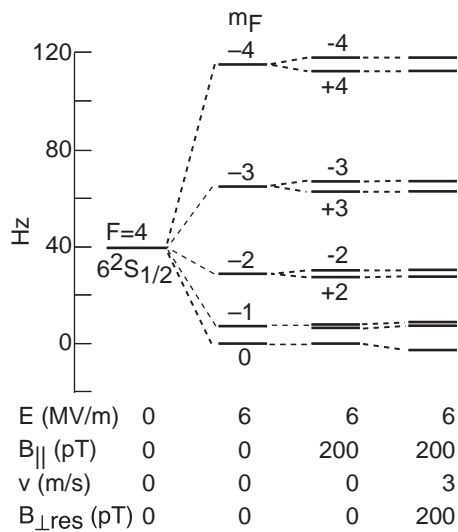}
\caption{\label{energy_levels}
Electric field quantized energy levels of the cesium ground state $6^2S_{1/2}, F = 4$, calculated from Eq. \ref{shift}. The conditions for the experiment reported here are represented by the rightmost column where the 3 m/s velocity results in a motional magnetic field of 200 pT. For comparison, the $\approx $70 ms transit time of the slow atoms through the electric field results in a transit time broadening of about 14 Hz.}
\end{figure}

A fountain e-EDM experiment can use two potent methods, not generally available to thermal atomic beam experiments, to suppress the motional magnetic field effect: atom-by-atom cancellation of the net beam velocity by the rise and fall of the slowly moving atoms under gravity, and electric field quantization. Using electric field quantization, no static magnetic field is needed because the electric field lifts the degeneracy of states of different $|m_F|$ (Fig. \ref{energy_levels}), and energy shifts due to the motional magnetic field are absent to first order \cite{player70}. 

Electric field quantization was first used in an e-EDM experiment by Player and Sandars \cite{player70} on the xenon $^3P_2$ metastable state which has a very large quadratic Stark effect. It was not possible to perform such an experiment on an alkali atom ground state because the alkali tensor polarizabilites are too small to lift the $m_F$ state degeneracy  past the several hundred Hz transit time broadening of a practical thermal atomic beam. But a fountain experiment can have a transit time broadening of one Hz, allowing tensor Stark splittings for heavy alkali atoms to be much larger than the transit time broadening. And even a beam of slow Cs atoms can be used. 

The incentive for pursuing this approach to improving the e-EDM limit is that it greatly suppresses the motional magnetic field systematic while preserving the desirable features of thermal atomic beams. These features include a simple and well understood system on which to experiment; experiments done in free space; the knowledge gained from thermal beam experiments; and the fruits of years of development of Cs fountain atomic clocks.

This paper describes an e-EDM experiment that is a
prototype for a Cs fountain experiment intended to reach a sensitivity of $2 \times10^{-50}$ C-m ($1.3 \times 10^{-29}$ e-cm), about two orders of magnitude below that of recent experiments \cite{regan02, hudson02, abdullah90, murthy89}. The present experiment demonstrates electric field quantization (with average magnetic fields below 200 pT); state preparation, transport and detection in magnetic and electric field-free regions; and separated oscillatory field type resonances between states with energy separations comparable to the transit time broadening. 
%
%
\section{EXPERIMENT}
\subsection {Electric Field Quantization \label{quant}}
 
In electric-field quantization, energy shifts due to the motional magnetic field are absent to first order \cite{player70}. The energy shift $W(m_F)$ of an $F = 4$, $m_F \ne 1$ sublevel in a strong electric field and with weak residual magnetic fields (Fig.~\ref{energy_levels}), and with the quantization axis defined by the electric field direction  is given by
\begin{eqnarray}\label{shift}
\frac{W(m_F)}{h} & = &\epsilon E^2 m_F^2 + g \mu B_{||} m_F \nonumber \\
& &  + K_1 \frac{(g\mu)^2 B_{\perp}^2}{\epsilon E^2} -K_2 \frac{(g\mu)^3 B_{\perp}^2 B_{||}}{(\epsilon E^2)^2}\\
& & -\frac{d_e R m_F E}{4h} + \mbox{higher order terms,}\nonumber
\end{eqnarray} 
where 
$\epsilon = -3 \alpha_T/56$, and
$\alpha_T \approx -3.5 \times 10^{-12}$ HzV$^{-2}$m$^2$ is the tensor polarizability of the $F = 4$, $m_F$ states \cite{gould69, ospelkaus03}, and $g\mu \approx 3.5 \times 10^9$ Hz/T, and $B_{||}$ is the component of magnetic field parallel to $\textbf{E}$, and $B_{\perp}$ is the component of magnetic field perpendicular to $\textbf{v}$ and to $\textbf{E}$, and $d_e$ is the e-EDM, $R$ is the enhancement factor, $h$ is Planck's constant, and $K_1$ and $K_2$ are given by
\begin{eqnarray}\label{coeff}
K_1(m_F) & = & \frac{m_F^2 +20}{2(4m_F^2 -1)} \\
K_2(m_F)& = & \frac{81 m_F}{2(4m_F^2 -1)^2}. \nonumber
\end{eqnarray}

Note that $B_{\perp}$ includes both $B_{\rm mot}$ and any static residual field $B_{\perp \rm res}$. The leading motional systematic effect $W_{\rm sys}(m_F)$ is then generated from the term in Eq. \ref{shift} that is proportional to $K_2$,  
\begin{eqnarray}\label{sys}
\frac{W_{\rm sys}(m_F)}{h} = -2K_2(m_F) \frac{(g\mu)^3 B_{\perp \rm res} B_{\rm mot} B_{||}}{(\epsilon E^2)^2}.
\end{eqnarray}
Here $B_{\rm mot}$ is found by Eq. \ref{mot} and $B_{\perp \rm res}$ is taken to be parallel to $B_{\rm mot}$. This term is odd in $E$ (through $B_{\rm mot}$) and odd in $m_F$ (through $K_2$) and thus mimics an EDM. This term can be suppressed, however, by making $E$ and $m_F$ large and by making $v$, $B_{\perp \rm res}$, and $B_{||}$ small. 

Under the conditions of this experiment ($E = 6$ MV/m, $v = 3$ m/s, and $B_{\rm mot} = B_{\perp \rm res} = B_{||} = 2 \times 10^{-10}$ T), reversing the electric field produces a shift of the $m_F = 4$ state equal to that produced by an e-EDM of $6 \times 10^{-46}$ C-m. In a fountain geometry, with a net residual velocity of 3 mm/s, the shift is equal in size to an e-EDM of $6 \times 10^{-49}$ C-m ($4 \times 10^{-28}$ e-cm) which is about a factor of four below the present experimental limit. Additional  reductions in $W_{\rm sys}$ are discussed in Section \ref{limit}.

\subsection{Apparatus}
\begin{figure}
\includegraphics [scale = 0.6] {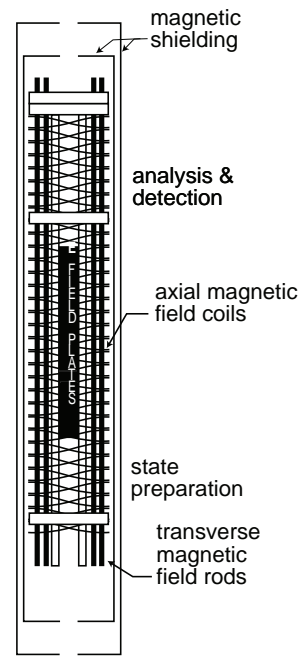} 
\caption{\label{interaction_region}Sketch of the interaction region looking along the direction of the electric field. The electric field plates are parallel to the plane of the page. Sixteen three-mm diameter copper rods, four of which are shown, were used to produce the nulling magnetic fields in the two directions transverse to the beam. Connections between the rods (not shown) were made at the top and bottom. The axial magnetic nulling field was produced by two solenoids wound with opposite pitch. The nulling coils were also used to produce the rotation and shifting pulses described in the text. The axial coils were used for the rotation pulse because there was less eddy current damping of the magnetic field in that direction from the electric field plate support structure. Vertical support rods and horizontal support plates are shown in white. The inner magnetic shields are also shown.}
\end{figure}
The rise and fall of atoms in a fountain results in an atom-by-atom cancellation of net velocity that greatly reduces the motional magnetic field systematic. Therefore to be able to test electric field quantization it was necessary to turn off the atom-by-atom velocity cancellation by increasing the atom's launch velocity to about 4.7 m/s so that the upward-traveling atoms did not turn around inside the electric field, but instead exited and were analyzed and detected above the electric field plates. This changed the fountain into a slow beam with an average upward velocity of about 3 m/s and a travel time between state preparation and analysis of about 150 ms (compared to about one second for a fountain). 

To the basic fountain apparatus, previously described in Ref. \cite{kalnins05, amini03, maddi99}, three sets of orthogonal magnetic field coils were added for nulling residual magnetic fields and for inducing transitions between states with different values of $m_F$. The field coils were surrounded by four magnetic shields --- two inside the type 304 stainless-steel vacuum chamber and two outside --- and by coils for demagnetizing the shields. The inner layers shielded against
magnetic fields from the vacuum chamber as well as from ferromagnetic 
seals on windows (needed for laser beams and to detect fluorescence) and on high-voltage feedthroughs. The windows and feedthroughs were mounted on ports that extended through the outer two layers of shielding.

Limitations of space prevented the openings in the shields (used for access to windows and high-voltage feedthroughs) from being surrounded by cylinders of additional shielding material and limited the space between the inner two shields; all this significantly reduced the shielding factor. The magnetic shields were fabricated from Co-Netic AA$^{TM}$ (Magnetic Shield Corp) and then annealed at 1120~$^{\rm o}$C: the outer shields in a hydrogen atmosphere but the inner shields in vacuum to avoid later outgassing of hydrogen into the vacuum system. 
Demagnetizing the shields in place left residual magnetic fields (even when the demagnetizing fields were smoothly ramped to zero) at points along the atoms' 
trajectory of typically a few nT. 

The residual magnetic fields were mapped 
in three orthogonal directions as a function of vertical 
position along the cesium atom's trajectory. This was done by applying and reversing additional magnetic fields from the three sets of orthogonal coils and measuring 
the frequency shift of transitions between $m_F$ states. We observed no hysteresis at additional fields of one $\mu$T.
Once the fields were mapped, waveform generators were programmed to deliver
time-dependent currents to the coils so that a local magnetic field null
was produced around the atom packet that followed the packet as it
traveled.  

Local maxima in the residual
magnetic field of about $3\,{\rm nT}$ were caused by 
magnetic fields entering through openings in the magnetic shields. The time dependent local nulling reduced the
fields experienced by the atoms to under $200\,{\rm pT}$ limited, most notably, by  the large gradients 
in the residual magnetic field. During data acquisition, the residual field was remeasured and the nulling recalibrated about once every 40 minutes.

Our legacy laser system was overmatched 
by the experimental requirements of trapping, launching, cooling,
state preparation, analysis, and detection --- all done with a single diode laser plus diode laser repumping.  Because of the weakness of this system and the defocusing of the atoms at the entrance and exit of the electric field,
only about 100 atoms were detected per launch.  

\subsection{State Preparation in a Field-Free Region}
\begin{figure}
\includegraphics [scale = 0.625] {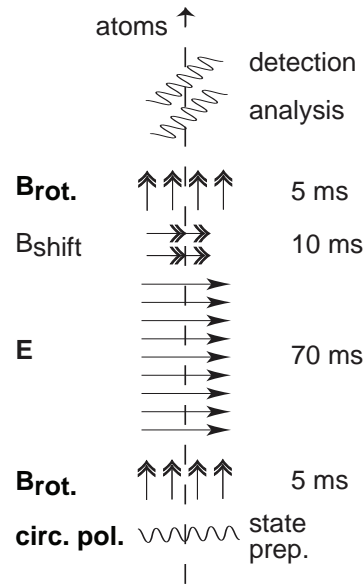} 
\caption{\label{MG_schematic} Schematic of the electric, magnetic, and optical fields. The wavy lines represent laser beams and the arrows represent electric and magnetic fields. Also shown are the time intervals during which the atoms experience the electric and pulsed magnetic fields. Drift times through free space are not shown. Quantities in bold are reversed in the course of the experiment. All magnetic field pulses are generated by coils that surround the entire region shown in the figure.  Because one packet of atoms travels upward through the apparatus at a time, all of the atoms in a packet experience the same fields. The quantization axis is parallel to the electric field and to the direction of the laser light used to prepare the initial state. The initial state is changed between $m_F = +4$ and $m_F = -4$ by changing the direction of circular polarization of the laser light used to prepare the state.  
 }
\end{figure}

After launching from the fountain's magneto-optical trap, and before entering the electric field, the packet
of cesium atoms enters the magnetically shielded and nulled region 
where the magnetic field affecting the atoms was measured
to be less than $200\,{\rm pT}$ and where all of the operations displayed in Fig.~\ref{MG_schematic} are performed.  In this essentially residual-field free region
atoms are prepared in the $F = 4, m_F = +4$ (or $m_F = -4$) state by
optical pumping to the $6^2P_{3/2}$, $F = 4$ level with circularly polarized light.
For the experiment to work, the optically-pumped atoms must remain 
in the $m_F = 4$ (or $m_F = -4$) state 
until they reach the electric field that will lift the $|m_F|$ degeneracy.
Because the residual magnetic field, $B_{\perp \rm res}$ perpendicular to the laser (and the electric field) was very small,
there was only a small (but detectable) mixing of the $m_F$ states.
There is similarly only a small (but detectable) mixing of the $m_F$ states 
due to $B_{\perp \rm res}$ throughout the region shown in Fig.~\ref{MG_schematic}

\begin{figure}
\includegraphics [scale = 0.625] {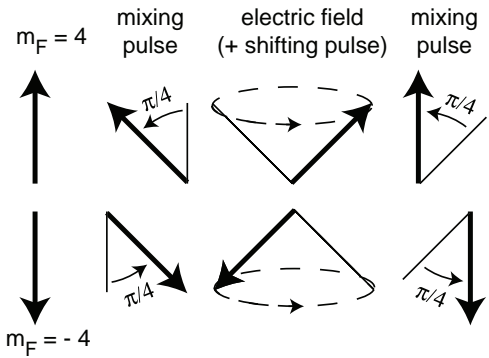} 
\caption{\label{rotation} Vector diagram of the state evolution. The upper row is for the initial $m_F = +4$ state and the lower row for the initial $m_F = -4$ state. In each case there is an initial and final $\approx \pi /4$ rotation pulse, which with the right amount of state precession in the electric field and if necessary in a shifting field, restores the atom to its original state. 
 }
\end{figure}

\subsection{Transitions Between Electric-Field Quantized States}
After state preparation, and while the atoms are still in the residual-field free region, a coherent superposition of $m_F$ states is generated by a 5 ms ``rotation'' magnetic field pulse parallel to the atomic velocity (see Fig.~\ref{MG_schematic}). The pulse amplitude is chosen to 
rotate the initial $m_F = 4$ state vector by an angle of
$\approx \pi/4$  (see Fig.~\ref{rotation}). 
The atoms then enter the electric field where each
$m_F$ state in the superposition gains a phase proportional to its energy ($\epsilon E^2 m_F^2$) in the electric field and to the time spent in the field.
The electric field of $\approx 6\,{\rm MV/m}$
is tuned so that the effect of passing through
the electric field is to rotate any initial
state vector by an angle of $\pi$ radians about the electric field axis (Fig.~\ref{rotation}). 

After exiting the electric field, a $10\,{\rm ms}$ pulse of magnetic field (shifting pulse)
parallel to the electric field direction is applied.  By varying the
magnitude of this ``shifting" magnetic pulse we can rotate the atomic state vector about the electric field axis.

A second 5 ms ``rotation'' magnetic field pulse parallel to the atomic velocity is applied to complete the  transition sequence, similar to the Ramsey separated oscillatory field method (Fig.~\ref{MG_schematic}). 
When there is no shifting pulse (and no e-EDM) the final state is $m_F = +4$.  
Finally, the percentage of the atoms that remained in states with $|m_F|= 4$ is measured as described in Section \ref{detection}.
\begin{figure}
\includegraphics  {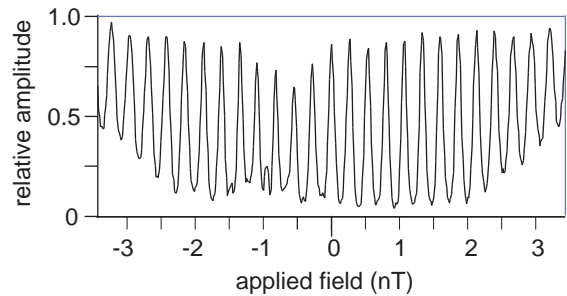}
\caption{\label{scan}
The detected sum of the population in $m_F = + 4$ and $m_F = - 4$ as a function
of the amplitude of a static magnetic field in the direction of the quantization axis. For this plot the full width of the resonances is set by the $90\,{\rm ms}$
transit time of the atoms from state selection to analysis. The loss of contrast
near $-0.7\,{\rm nT}$ is consistent with a $0.3\,{\rm nT}$ remnant
magnetic field perpendicular to the electric field. 
 } 
\end{figure}

The probability that the final state
is a state with $|m_F| = 4$ is periodic (with period $2\pi$)
in the state vector rotation about the electric field axis. 
The rotation about the electric field axis can be scanned by varying a weak static magnetic field (applied for the entire 90 ms) in the direction of the electric field axis (with or without the electric field). This produces
data such as that shown in Fig.~\ref{scan}.  
As we show later, an EDM manifests itself as a horizontal translation of
the resonances that is odd in the sign of the electric field.

\subsection{Transition Lineshape}
Take any initial state $\Psi$
within a hyperfine level, apply
any perturbation that only mixes states within
the level, then apply a shifting pulse that rotates the result by
an angle $\phi$, and compute the projection of the result upon
some specified state $\Psi'$ within the level; the
observable
$$T(\phi) = \big|\big<\Psi\big|\Psi'\big>\big|^2$$
is necessarily a real function of $\phi$ of period $2\pi$.  Such an observable
therefore has a standard Fourier series expansion
$$T(\phi) = \sum c_m e^{im\phi}$$
with complex coefficients
$$c_m = {1\over 2\pi}\int_{-\pi}^\pi e^{-im\phi}T(\phi)\,d\phi\ ;$$
for a hyperfine level of total spin $F$
only the coefficients $c_m$ for $|m|\le 2F$ can be nonzero.

An e-EDM rotates the state vector along the same axis as does the
shifting pulse, assuming the electric and the shifting pulse fields are parallel;
the lineshape function therefore changes to
$$T(\phi) = \sum_{m=-F}^F c_m e^{i(\phi+\eta)m}\ ,$$
where the new angle is
$$ Rd_eE\tau_E/4\hbar,$$
where $\tau_E$ is the time the atoms spend in the electric field.
An e-EDM therefore translates the lineshape without distortion;
the basic idea behind the data analysis is therefore to look
for a translation of a lineshape that reverses when the electric field
reverses, but not when the initial states $m_F = +4$ and $-4$ are
exchanged or when the common polarity of the rotation pulses is reversed.
It is useful that every detail of the actual experimental lineshape
does not have to be understood to extract a value of an e-EDM from
its translation.  

\begin{figure}
\includegraphics [scale = 0.625] {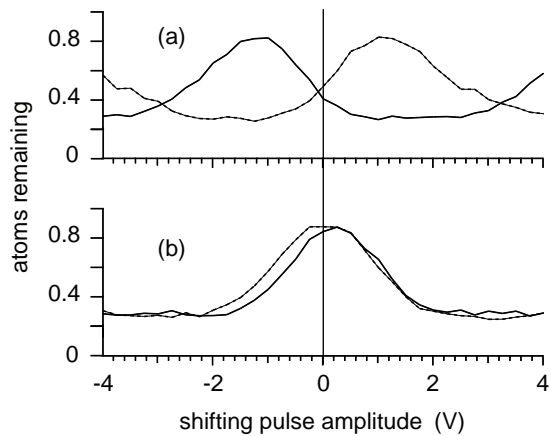} 
\caption{\label{critical_field} Scan of shifting pulse for initial states $m_F = 4$ (solid line) and $m_F = -4$ (broken line) for (a) the electric field set at $98\%$ of the field needed to produce a rotation of $\pi$ radians and (b) the electric field set to produce a rotation of $\pi$ radians.
 }
\end{figure}

While it is not the detail of a lineshape, but merely its translation,
that is the signature of an e-EDM, experimentally it is helpful to have
that lineshape as simple as possible.
Given an initial hyperfine state $\big|FM\big>$,
a time-dependent electric field $E$ parallel to the quantization axis
introduces a phase $e^{-iM^2\theta}$, where
$$\theta = \epsilon \int E^2(t)\,dt $$
and $\epsilon$ was defined in Section \ref{quant}. 
A rotation of the state vector about the axis by an angle
$\phi$ would introduce instead a phase $e^{-iM\phi}$;
a rotation by $\phi = \pi$ therefore
introduces a phase $+1$ if $M$ is even and $-1$ if $M$ is odd.

Precisely the same phases are introduced by the electric field
if we set $\theta = \pi$, whereupon the generally complicated effect of an
electric field on an arbitrary state within the hyperfine level reduces
to a simple rotation of that state about the field axis by an angle $\pi$.
Under this condition, the lineshape produced by varying the rotation
of the state vector (by scanning the shift field)
when the electric field is on, is identical with the lineshape produced by
varying the rotation with when the electric field off, except that the
lineshape is translated in rotation angle by $\pi$; in this sense
the electric field then does not distort, but merely translates, the lineshape.

The value $\theta$ can be set very close to~$\pi$  even though
the cesium tensor polarizability, and hence the parameter
$\epsilon$, is known to no better than roughly~6\%~\cite{gould69,ospelkaus03}. When $\theta$
departs from $\pi$, the lineshape not only distorts, but translates,
and this translation is in opposite directions for the initial states
$M=+4$ and $-4$, as shown in Fig.'s~\ref{rotation} and \ref{critical_field}; only for $\theta = \pi$ do the
lineshapes for the different initial states superimpose.
In our apparatus
the condition $\theta = \pi$ is met for an electric
field of $\approx 6\,{\rm MV/m}$;  our plates would not sustain the fields required to explore values
of higher integer multiples of~$\pi$.

\subsection{State Analysis and Detection in a Field-Free Region \label{detection}}
The fraction of atoms that remain in states with $|m_F|= 4$ is measured
by transferring the population in states with $|m_F|\ne 4$ into the 
empty $F= 3$ hyperfine level and then counting the atoms remaining in the $F=4$ level. For normalization, the atoms in the $F=3$ level are pumped back into the $F=4$ level and
all of the atoms are detected.

The transfer of states with $|m_F|\ne 4$ into the empty $F= 3$ hyperfine level is accomplished using
light polarized parallel to the electric field. This light excites all but states with $|m_F|=4$
into the $6^2P_{3/2}$, $F=3$ level, which decays 3/4 of the time
to the ground state $F=3$ hyperfine level. The remaining 1/4
of the time the atom returns to the ground state $F=4$ hyperfine level.  After 
many cycles, the population of states with $F = 4, |m_F|=4$ states is the sum
of the original populations, plus $20\%$
of the population originally in states with $|m_F|=3$, plus a smaller percentage
of the population originally in other $F = 4, m_F$ states.

The atoms remaining in the $F=4$ hyperfine level
are detected by exciting the cycling transition
$6^2S_{1/2}$, $F=4$ to $6^2P_{3/2}$, $F=5$ and collecting the
fluorescence radiation into a photomultiplier. The atoms in the $F=3$ hyperfine level
are then pumped back into the $F=4$ hyperfine level and all of atoms detected
by again exciting the cycling transition.

By the time the atoms have reached the detection region, they have spread longitudinally to many times the width of the viewing region of the detector. A millimeter-sized region of passing atoms are detected and then normalized by chopping between the two laser beams and synchronously switching the output of the detector into counters for signal and normalization.
\begin{figure}
\includegraphics {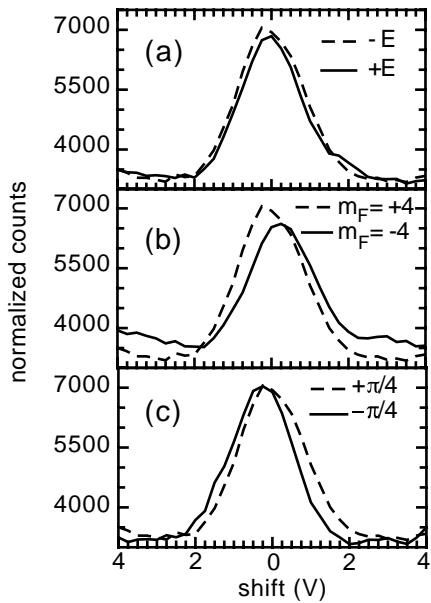}
\caption{\label{resonance}
The $|m_F| = 4$ population measured as a function
of the amplitude of the shift pulse (the conversion is $1\,{\rm V}\approx
100\,{\rm pT})$.  The resonance is periodic in the shift pulse
amplitude and slightly more than one period is plotted.
Shown are the effects on the resonance
position and shape of: (a) a reversal of the electric field; (b)
a change of the initial state between $+4$ and $-4$; and (c)
a change in the common polarity of the rotations. For ease of reference,
the broken line shows a common condition of $E$, of $m_F = +4$, and
the polarity of the rotations.  An e-EDM (or systematic error) of
$4\times 10^{-43}$ C-m ($2.5\times 10^{-22}$ e-cm)
would produce a resonance shift of about 0.1 V.
}
\end{figure} 

\subsection{Results}
Resonance shapes were measured for the two electric field polarities,
for the initial states $m_F=4$ and $m_F = -4$, and for both common
polarities of the 5 ms rotation pulses -- a total of eight combinations.
A signature of an e-EDM is a shift in the $m_F$ state energy (a change in the accumulated phase due to the atom's traversing the electric field plates) that is odd under a reversal of the electric field polarity, odd under a change in initial state from $m_F = +4$ to $m_F = -4$ and even under a reversal of the common polarity of the rotation pulses. Reversing the electric field cancels out terms in Eq.~\ref{shift} that are independent of $E$, or that are even in $E$, such as $B_\parallel$ and the tensor Stark shift (the $\epsilon E^2 m_F^2$ term in Eq ~\ref{shift}). Reversing the sign of $m_F$ for the initial state cancels terms that are even in $m_F$ and therefore cancels the effects of an incomplete reversal of the electric field and cancels the term in $K_1$ in Eq's ~\ref{shift} and \ref{coeff}. Any difference in the centroids for $m_F = +4$ and -4 due to a failure to set the magnitude of the electric field to produce a rotation of precisely $\pi$ radians (see Fig.\ref{critical_field}) also cancels.

To search for an e-EDM, the fraction of Cs atoms remaining in the $|m_F =4|$ state was measured as a function of the amplitude of the shifting pulse (see Fig. \ref{resonance}) for each of the eight combinations of electric field polarity, sign of initial $m_F = \pm 4$ state, and common polarity of the rotation pulses. If the scan of surviving $|m_F|$ state fraction as a function of shifting pulse amplitude is free from distortions that might change under some combination of reversals, it is only necessary to measure the surviving $|m_F|$ state fraction at a few values of the shifting pulse (generally where the slope is largest) and observe any change in the fraction of atoms detected in the $|m_F| =4$ state upon reversal of the sign of the electric field and/or the initial state. This is the traditional way to take e-EDM data because it allows one to make frequent reversals and so cancel out (residual) magnetic field drifts and other drifts. 

However the scans in Fig. \ref{resonance} may deviate from the sinusoids that would be predicted for a two level system because the nine $m_F$ states in the $F =4$ hyperfine level are all coupled by the rotation pulses, by the motional magnetic field, and by residual perpendicular magnetic fields. Therefore, the surviving $|m_F|$ state fraction was mapped as a function of shifting pulse amplitude as shown in Fig. \ref{resonance}. Unfortunately, a set of eight maps took 40 minutes, leaving the measurement vulnerable to slow drifts in the magnetic field whose effects could otherwise be cancelled by frequent reversals of the electric field.

Eighteen sets of the eight combinations of reversals yielded a total of about $5 \times 10^5$ detected atoms.  The result is an e-EDM limit of $-0.7 \pm 2.2 \times 10^{-43}$ C-m ($-0.5 \pm 1.4 \times10^{-22}$ e-cm) where the value in parenthesis is the statistical uncertainty at the $1 \sigma$ level. At this level of precision the residual motional magnetic field systematic (Section \ref{quant}) is not a factor in the measurement. 

\section{IMPROVING THE e-EDM LIMIT \label{limit}}
In this section the possibility of improving the e-EDM limit in a cesium fountain experiment, with electric field quantization to suppress the motional magnetic field systematic, multiple quantum transitions and electrostatic focusing to improve sensitivity, and high resistivity materials to reduce magnetic Johnson noise, is considered.

In an apparatus where a rise and subsequent fall of atoms reduces the time-averaged velocity to $< 3$ mm/s, the motional magnetic field effect is immediately reduced by a factor of $10^3$ compared to the present experiment. The residual velocity is set by a possible transverse drift of the beam or by timing uncertainties in the mixing pulses. 
An earlier experiment using the present fountain measured the change in longitudinal velocity of Cs atoms entering an electric field as a function of electric field strength to determine the Cs static polarizability \cite{amini03}.

In an improved apparatus, the static residual magnetic fields, $B_{\perp \rm res}$ and $B_{||}$  might each be reduced a factor of ten or more to $< 2 \times 10^{-11}$ T through improved shielding design, using thicker shields, adding additional layers of shielding, and using external coils for active shielding. Combined with the fountain geometry, this would reduce the motional magnetic field systematic $W_{\rm sys}$ compared to the present experiment by a factor of $10^{5}$ (See Eq. \ref{sys}). 

Increasing the electric field from $\approx 6$ MV/m to 13.5 MV/m would bring the total reduction in $W_{\rm sys}$ to about a factor of $10^6$. Reversing the electric field would then produce a shift of the $m_F = 4$ state equal to that produced by an e-EDM of $6 \times 10^{-52}$ C-m.
As in the present experiment, a possible systematic from incomplete reversal of the electric field is subtracted out by reversing the sign of the initial state from $m_F = +4$ to $m_F = -4$ and by monitoring the electric field plate voltages.

Many improvements to the experimental sensitivity are also possible. The fountain geometry would reduce transit time broadening to about 1 Hz. Using seven-quantum transitions $m_F = \pm 4 \leftrightarrow m_F = \mp 3$ would produce an additional factor of seven reduction in the transit time broadening compared to a single photon transition. The seven quantum transition appears feasible if the oscillatory fields or rotation pulses are applied while the atoms are in the electric field. 

Multiple quantum transitions with line narrowing using separated oscillatory fields have been observed in Tl \cite{gould76} and 
line narrowing effects have been observed in Cs \cite{xu99}. Increasing the electric field from $\approx 6$ MV/m to 13.6 MV/m would also increase the e-EDM sensitivity. With the fountain, seven-quantum transitions, and the high electric field, about $2 \times 10^{14}$ detected atoms would be needed to reach an e-EDM sensitivity of $2 \times 10^{-50}$ C-m (about a factor of 100 below the present experimental limit).

The time needed to reach this statistical sensitivity depends upon the flux and temperature of the cesium atoms, their survival in the fountain, the transition probability, and the detection efficiency. For a real experiment, time for systematic tests, magnetic field nulling, beam tuning, etc., as well as maintenance and repairs, must be added. State selective detection efficiency can be 80\% and the seven-quantum transition probability is calculated to be close to 90\%.  Cesium atom fluxes of $>1\times10^9$ s$^{-1}$  have been launched and cooled to 1.5 $\mu$K or lower \cite{legre98,treutlein01}. 

To have all or most of these atoms return, it is not sufficient that the atoms be cold and the electric-field plate gap be large. It is also necessary to focus the atoms to counter the defocusing effect of the electric-field gradient at the entrance of the electric-field plates \cite{maddi99} and the heating of the atoms (by $\approx 2 \mu$K) due to the optical pumping into the $m_F = \pm4$ state. Electrostatic focusing does not introduce any magnetic fields and focuses all of the $m_F$ states identically because the tensor polarizabilities are much smaller than the Cs ground state scaler polarizability. 

An electrostatic lens triplet, designed from first principles, has been used with the present Cs fountain to produce focused beams and parallel beams of Cs atoms\cite{kalnins05}. 
Simulations \cite{kalnins03} show that a combination of an electrostatic triplet plus an electrostatic doublet can compensate for beam heating and defocusing. Focused into a near parallel beam, nearly 100\% of the atoms entering a pair of electric field plates with a 10 mm gap spacing and 13.5 MV/m field would return to be detected.
In addition, the trajectory of the fountain and hence the transverse drift of the atoms would be controlled by focusing lenses. 

To significantly improve the e-EDM limit it is also necessary to reduce the magnetic Johnson noise \cite{munger05}. This generally means substituting higher resistivity materials for the metals traditionally used in the electric field plates, the vacuum chamber, and possibly the innermost magnetic shield. Electric field plates may be made from soda lime glass (such as Corning type 0080), which when heated to about 150 $^o$C become sufficiently conductive. Glass electric field plates will sustain higher electric fields than metal plates of the same dimensions, making it easier to reach an electric field of 13.5 MV/m with a ten mm gap spacing. Heated glass electrodes have previously been built and used for polarizability measurements on Tl and Cs thermal beams \cite{gould76, marrus69}.
A metal vacuum chamber may be replaced by a (mostly) glass chamber and the innermost magnetic shield can be made of ferrite \cite{kornack07}. 

To turn these possible improvements in systematic suppression, e-EDM sensitivity, and magnetic noise reduction into real experimental gains, many experimental details, not discussed here, must also be worked out.

\section{CONCLUSION}
In a proof-of-principle experiment, electric field quantization of a slow beam of cesium atoms has been achieved in an electric field of 6 MV/m with the magnetic field seen by the atoms reduced to less than 200 pT. The atoms are optically pumped, transported, undergo transitions induced with separated pulsed fields, and are analyzed and detected --- all in regions free of applied static magnetic and electric fields. Although the present experiment was limited (mostly) by our laser system, these techniques may be used to lower the e-EDM limit by two orders of magnitude in a full scale cesium fountain  experiment. Such an experiment is being planned by two of us (H.G and C.T.M.).
\section*{ACKNOWLEDGMENTS}

We thank Timothy Dinneen for assistance in constructing the original fountain and Douglas McColm for early theoretical guidance. We gratefully acknowledge support of a NIST Precision Measurements Grant and support from the NASA Fundamental Physics Program,  Office of Biological and Physical Research. The Lawrence Berkeley National Laboratory is operated for the U.S. DOE under Contract  No. DE-AC02-05CH11231.

\bibliography{EDMbib}
\end{document}